# Chandra Detection of the Circumnuclear Molecular Torus of the Compton Thick AGN in NGC 5643

*G. Fabbiano[1], A. Paggi[2,3,4], A. Siemiginowska[1], M. Elvis[1]*

1. Center for Astrophysics | Harvard & Smithsonian, 60 Garden St. Cambridge MA 02138, USA
2. Dipartimento di Fisica, Universita' degli Studi di Torino, via Pietro Giuria 1, I-10125 Torino, Italy
3. Istituto Nazionale di Fisica Nucleare, Sezione di Torino, via Pietro Giuria 1, 10125 Torino, Italy
4. INAF-Osservatorio Astrofisico di Torino, via Osservatorio 20, 10025 Pino Torinese, Italy

## *ABSTRACT*

We report a clumpy elongated feature found with deep *Chandra* ACIS high-resolution imaging of the Fe Kα line emission in the nuclear region of the Compton Thick Active Galactic Nucleus (CT AGN) galaxy NGC 5643. This feature extends for ~65 pc N-S. No corresponding feature is seen in the 3.0-6.0 keV continuum. The Fe Kα feature is spatially consistent with the N-S elongation found in the CO(2-1) high resolution imaging with *ALMA* (Alonso-Herrero et al 2018), but slightly more extended than the rotating molecular disk of r=26 pc indicated by the kinematics of the CO(2-1) line. The *Chandra* detection of a corresponding N-S structure in the neutral Fe Kα line, would argue for both CO and Fe Kα emission originating from the obscuring torus.

## 1. Introduction

NGC 5643 is a face-on barred southern galaxy (J2000 RA = 14h32m40.743s, Dec. = -44d10m27.86s) hosting a Compton Thick Active Galactic Nucleus (CT AGN) with nuclear $N_H>10^{24}$ cm$^{-2}$ (Risaliti et al 1999). This AGN has a double-sided ionization cone in the direction of a double radio jet originating from the nucleus (Morris et al 1985; Schmitt et al 1994; Cresci et al 2015). NGC 5643 has been observed with several X-ray observatories to both study its active nucleus, and more recently its population of galactic sources, including an ultraluminous X-ray source (e.g., Guainazzi et al 2004; Matt et al 2013; Annuar et al 2015). Recent *ALMA* observations (Alonso-Herrero et al 2018) have led to the discovery of a nuclear rotating obscuring disk with 26 pc diameter in this AGN (0.3", given the assumed distance of 16.9 Mpc; for ease of comparison, we assume the same distance in this paper). This molecular disk is oriented in the north-south direction, perpendicular to the ionization cone. This disk could be related to the obscuring torus of the standard AGN model, responsible for the nuclear obscuration and for the collimation of the ionization cone.

There have been reports with *Chandra* imaging of features possibly connected with the obscuring torus in AGNs, but never in conjunction with an independent observational validation of the existence of this torus. Extended clumpy emission in the 3-6 keV continuum and in the Fe K$\alpha$ line emission (both neutral and ionized) have been discovered in the nearby CT AGN NGC 4945 with *Chandra* (Marinucci et al 2012; 2017) and related to the obscuring torus. More recently, we have discovered clumpy Fe K$\alpha$ emission within ~30pc of the nucleus of the CT AGN ESO 428-G014), and suggested that this emission originate from fluorescent emission of the nuclear photons interacting with thick obscuring clouds (Fabbiano et al 2018c. The detection of the circumnuclear disk in CO(2-1) emission with *ALMA* in NGC 5643, and the existence of a deep *Chandra* ACIS dataset for this galaxy, provide a unique opportunity for exploring both millimeter and X-ray diagnostics of a molecular optical disk torus in CT AGNs.

The only report in the literature of the morphology of the hard continuum and Fe K$\alpha$ nuclear emission in NGC 5643 is consistent with a *Chandra* point source (Annuar et al 2015). However, the *Chandra* data used by Annuar et al (ObsID 5636) were the result of a very short (7.6 ks), 4' off-axis ACIS observation, where the PSF would be significantly wider (90% encircled energy fraction > 3", Chandra POG: http://cxc.harvard.edu/proposer/POG/html/chap4.html). Subsequently, NGC 5643 has been observed twice with ACIS, with the AGN at the aim point of the *Chandra* telescope, resulting on a total exposure of 113.4 ks. In this paper, we report the results of the analysis of this deep *Chandra* data set, where the nucleus is observed at the best possible angular resolution. This analysis led to the discovery of N-S extended emission of the nuclear Fe K$\alpha$ emission of NGC 5643.

## 2. Data and Analysis

Table 1 summarizes the two observations of NGC 5643 we used for this study. These are the only two observations in the *Chandra* archive that place the nucleus of NGC 5643 at the aim point, resulting in the best possible angular resolution. The data and basic reduction and analysis procedures used here follow closely those used in our previous work (e.g., Fabbiano et al 2018b, where we report the spatial analysis of the nucleus of the CT AGN ESO 428-G014).

Table 1. Observation Log

| ObsID | Instrument | $T_{exp}$(ks) | PI | Date |
| --- | --- | --- | --- | --- |
| 17031 | ACIS-S | 71.9 | Fabbiano | 2015-05-21 |
| 17664 | ACIS-S | 41.5 | Fabbiano | 2015-12-26 |

All the data sets were screened for high background times, processed to enable sub-pixel analysis and merged using standard *CIAO* (see Fruscione et al 2006) tools (http://cxc.cfa.harvard.edu/ciao/).

We use 1/8 sub-pixel binning for this analysis (the ACIS instrument pixel is 0''.492; see Fabbiano et al 2018 b, c for discussions on and similar applications of sub-pixel binning). For our analysis, we used *CIAO* 4.8 tools and the display application *DS9* (http://ds9.si.edu/site/Home.html), which is also packaged with *CIAO*. To highlight some of the features, we applied a Gaussian kernel smoothing to the images, as described in the figure captions.

The *Chandra* PSF was simulated using rays produced by the *Chandra Ray Tracer* (*ChaRT*) (http://cxc.harvard.edu/ciao/PSFs/chart2/) projected on the image plane by *MARX* (http://space.mit.edu/CXC/MARX/). We refer to Fabbiano et al 2018c for an in-depth discussion and validation of the shape of the *Chandra* PSF at the energies considered in this paper and using sub-pixel data.

We checked each individual file to ensure that the N-S extended 6.2-6.6 keV feature that we are reporting in this paper is not a byproduct of the PSF anomaly (http://cxc.harvard.edu/ciao/caveats/psf_artifact.html). Both the orientation of the N-S feature and the relative intensity of the extended versus central emission exclude this possibility. The artifact is expected to show a S-E direction in ObsID 17664 and a S-W direction in ObsID 17031, contrary to the N-S extension in the data. Moreover, the artifact may display ~6% of the source counts, much less than detected in the N-S extension (see below, Figure 3).

We merged the two observations in Table 1, using the several bright point sources detected in the ACIS field to cross-match the images, excluding the central source of NGC 5643. All these sources were consistent with being point-like after merging. The images from each ObsID and the merged image, all show clear evidence of extended emission from the ionization cone and a bright nuclear source. These features have different spectral signatures (as in the case of other CT AGNs, see Levenson et al 2006; Fabbiano et al 2018a). Fig. 1 shows the spectral count distribution of all the data connected with the extended and nuclear emission of NGC 5643 (no detected point sources are in the extraction area, so this is the Chandra spectrum of CT AGN emission). This spectrum shows soft emission-line dominated emission at energies below 2.5 keV, which is connected with the extended ionization cone (and will be discussed elsewhere). At higher energies, the spectrum is dominated by a featureless flat continuum and by the Fe Kα neutral emission line at 6.4 keV. As in our study of ESO 428-G014, we imaged separately hard continuum (3.0-6.0 keV) and Fe Kα emission (6.2-6.6 keV). These images are shown in Figs. 2 and 3, compared with the PSF appropriate for these energies and focal plane position. Fig. 2 shows that the 3.0-6.2 keV continuum source is somewhat extended (this will be discussed elsewhere), but with no pronounced directionality of the surface brightness. Instead the 6.2-6.6 keV image shows a marked N-S extension in the nuclear source (bottom right panel). This extent is also present in the same band, and with the same directionality, in each of the two separate observations (Fig. 3; left panels). This and the point-like appearance of the continuum emission peak exclude bad astrometry in the merging.

Fig. 3 shows a simulated PSF obtained by merging the two individually simulated PSF tailored to each an axis observation. This PSF is clearly centrally peaked, contrary to the observations. From the model PSF we calculate a ratio $R_{PSF}$=5 between the counts enclosed within a circle of 0".2 radius centered on the peak of the PSF (identified by a cross in Fig. 3), and an adjacent circle of same radius. A comparison with our longest on-axis observation (17031) gives a consistent ratio (given the large statistical error), when comparing the central circle with the one in a fairly low-surface brightness region at the left (to the east; in blue in Fig. 3). Instead, we obtain a ratio of $R_S$=1.6±0.5 when comparing the central circle with the one to the south, a ~7$\sigma$ lower than indicated by the PSF. The shorter on-axis observation (17664) also shows a N-S elongation, albeit with more limited statistics. Within statistics, the counts from all the regions shown in the merged images in Fig. 3 are consistent in the two separate observations. Using the merged image, we obtain, $R_N$=2.1±0.5 (a 5.8$\sigma$ from the PSF ratio), $R_S$=1.4±0.3 (12$\sigma$ discrepancy), showing a N-S extension at high statistical significance. Instead, $R_E$=4.2±1.8, consistent with the PSF. Similar conclusion are arrived to if instead of directly calculating the statistical errors from the data (using the square root of the counts, and propagating them as appropriate), we compare the ratios of the counts from a given region to the nuclear counts from the data with the results of 1000 simulations based on the PSF model normalized to the total number of counts in the extended nuclear region (Fig. 4). In these simulations the ratio for the E region is also consistent with that expected from a point source.

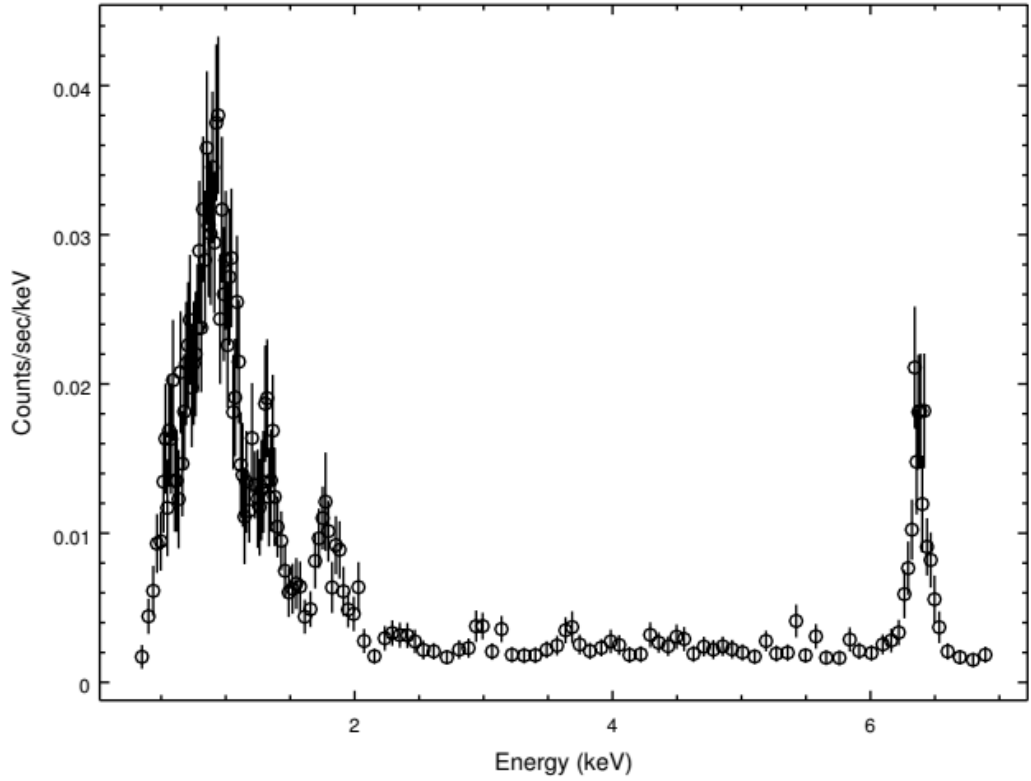

Fig. 1- Observed spectral distribution of the nuclear emission of NGC 5643. These counts were extracted from a ~30"× 20" box enclosing the nucleus and the entire extended emission of the ionization cone.

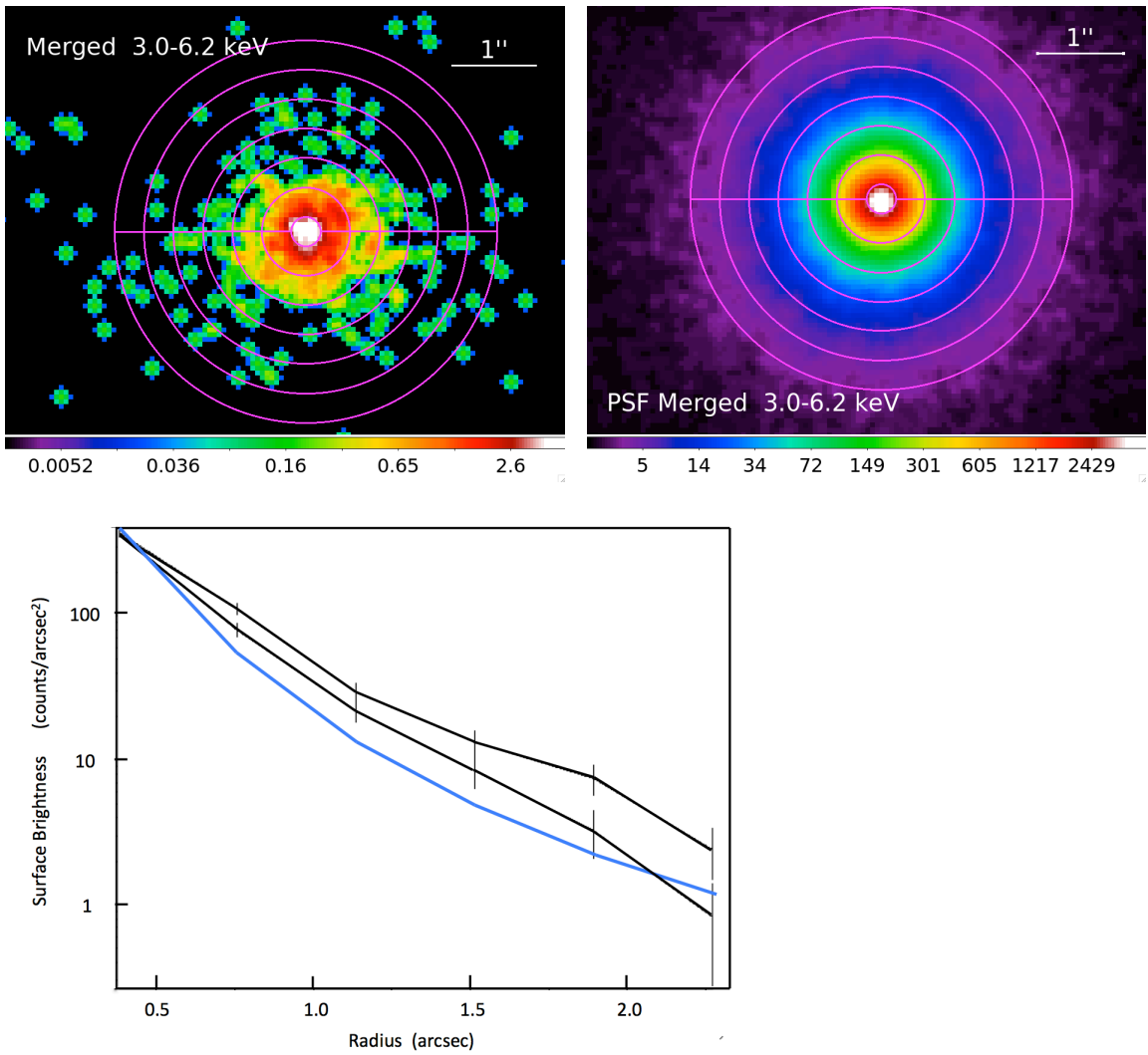

Fig. 2 – Top left: 1/8 pixel, (3.0-6.2 keV) image of the nuclear region, with a 2 pixel Gaussian smoothing applied for visibility. Top right: the corresponding model PSF. The intensity color scale (in counts/pixel) is given at the bottom of each panel. N at the top and E to the left of each panel. On both images are superimposed the extraction regions used to derive the radial profiles shown in the bottom panel: in blue is the PSF, in black the N and S hemispheres from the data.

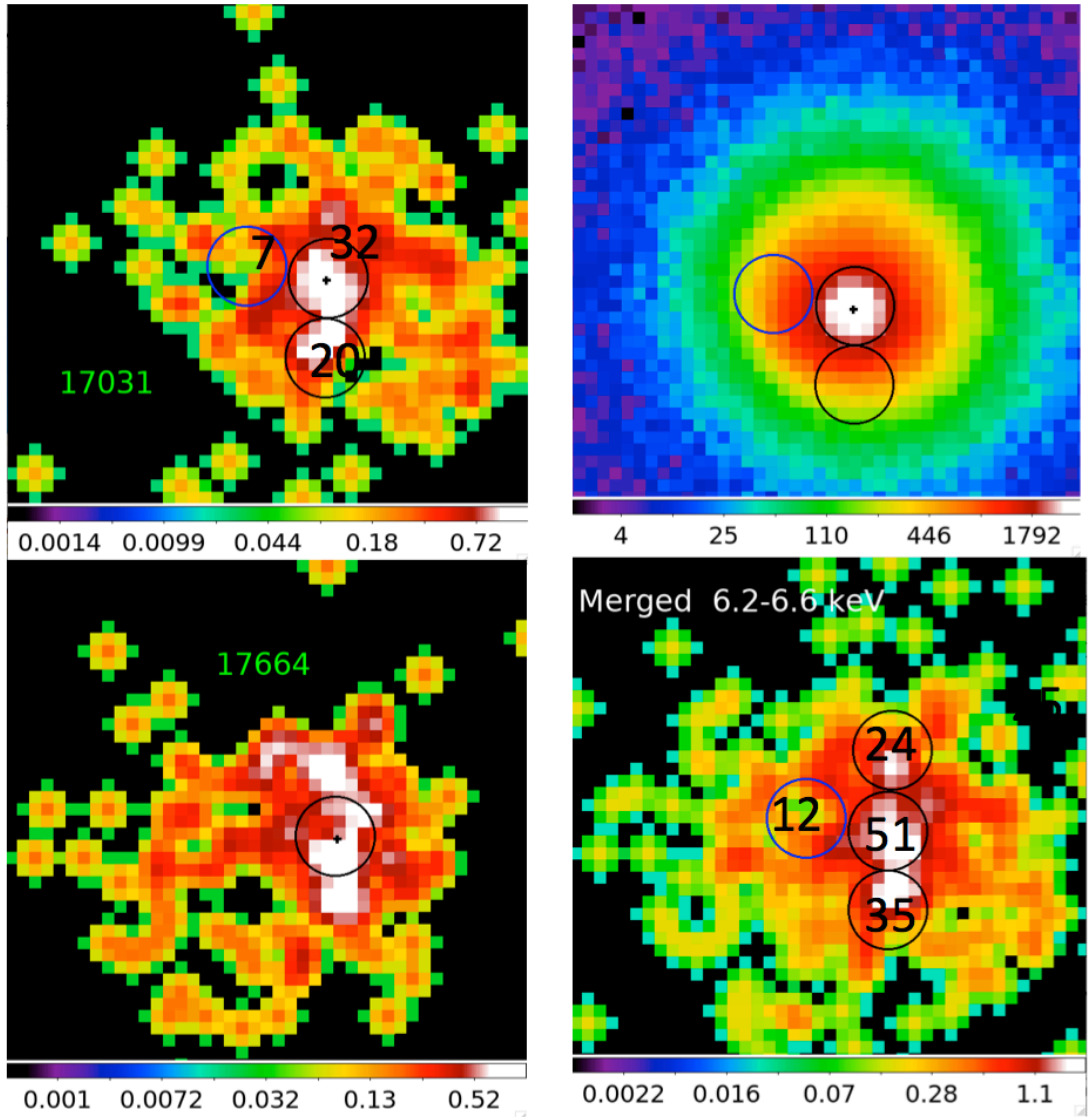

Fig. 3 – From the bottom panel, clockwise: the merged 6.2-6.6 keV image; the images of the two on-axis observations in the same spectral band, as indicated; and the 6.2-6.6 keV merged PSF. All images are binned in 1/8 pixel. The data have been smoothed with a 2 pixel Gaussian for visibility. The intensity color scale (in counts/pixel) is given at the bottom of each panel; these scales are different in each panel. The 0''.2 radius circular extraction areas, and counts (see text) are also plotted. N at the top and E to the left of each panel.

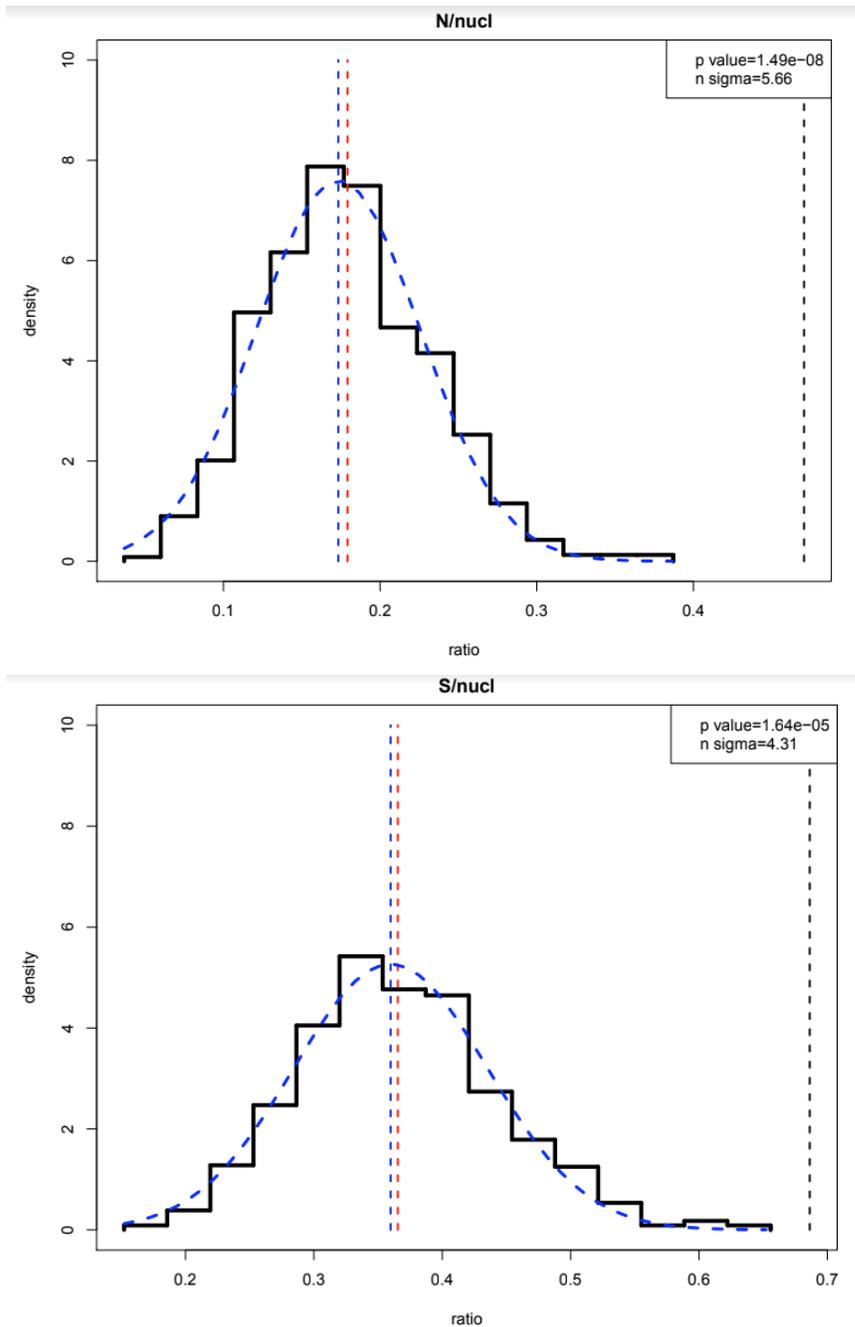

Fig. 4 – Probability density distributions of the simulated ratios from 1000 instances of a model PSF normalized to the total counts in the 6.2-6.6 keV extended nuclear region (solid line). Top: for the ratio of the region N of the nucleus in Fig. 3 relative to the nucleus. Bottom, same for the region S of the nucleus in Fig. 3. The Gaussian fits to the distributions are represented by the blue dashed lines, with a mean indicated with the vertical dashed blue line. The ratio in the psf model is indicated with the vertical red dashed line, and the observed ratio in the region is indicated with a vertical dashed black line. In the box are indicated the probability of observing such ratios given the fitted gaussian distribution and the equivalent sigmas.

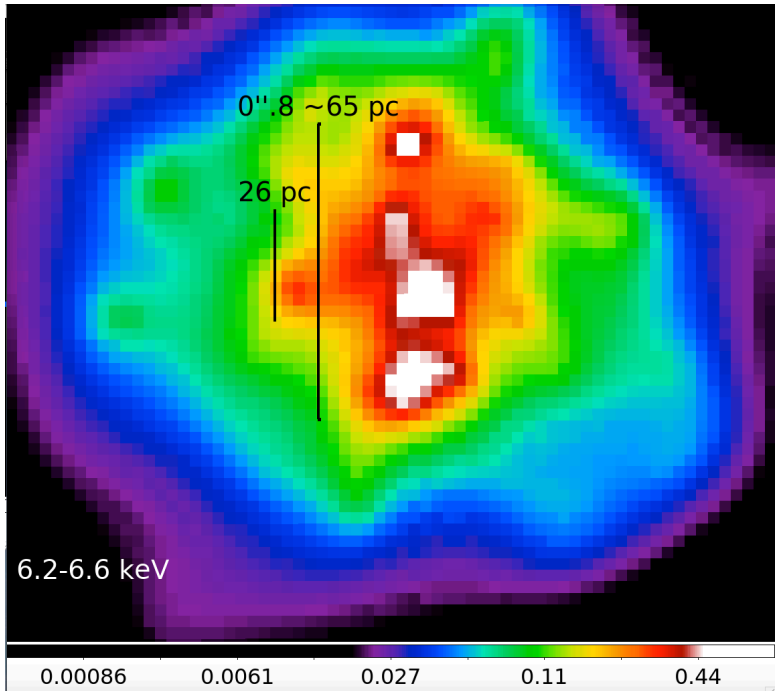

Fig. 5 – Left: Adaptively smoothed image of the 6.2-6.6 keV nuclear emission with 1/16 pixels (see text). We show also the size of the *ALMA* disk (26 pc, Alonso-Herrero et al 2018). The intensity color scale (in counts/pixel) is given at the bottom. N at the top and E to the left.

The above analysis shows that nuclear emission is more complex and extended in the N-S direction than expected from a single point source. This is confirmed by $\chi^2$ tests. We compared via $\chi^2$ data from N-S and E-W extraction strips consisting of three adjacent square boxes of 0".3 side, the central of which including the higher surface brightness nuclear region (see Figs. 2 and 3) with counts from similar regions extracted from the model PSF. Normalizing the PSF either to the counts in central 'nuclear' region, or to the total number of counts in the three bins, we obtain p-values P~0.02 - 0.05 for the E-W direction, suggesting that the count distribution could be marginally consistent with that of a point source. In the N-S direction instead we obtain p-values P << 0.001 in both cases, showing that the emission is significantly more extended than the PSF.

To get a better definition of the morphology of the N-S Fe K$\alpha$ structure, we have produced a 1/16 pixel image and processed with adaptive smoothing, using Gaussian kernels ranging from 0.5 to 10 pixels, 30 iterations, and 5 counts under the kernel. The result is shown in Fig. 5. The N-S feature extends for ~0.8", corresponding to 65 pc at the assumed distance of 16.9 Mpc.

**3. Discussion**

High-resolution imaging studies of *Chandra* ACIS observations, in different spectral bands, are starting to show the complexity of the inner (r~100 pc) circumnuclear regions of CT AGNs. NGC 5643, discussed here, is the first example where *ALMA* and *Chandra* observations can be used to give a complementary picture of the obscuring torus.

3.1 Comparison with other circumnuclear regions of CT AGNs studied with *Chandra*

In the nearby CT AGN NGC 4945, possibly the best-studied case so far (Marinucci et al 2012, 2017), a ~200 × 100 pc clumpy flattened distribution perpendicular to the main axis of the ionization cone was detected with *Chandra* in both hard continuum (~3-6 keV), 6.4 keV neutral Fe Kα, and 6.7 keV Fe XXV emission lines. In the CT AGN ESO 428-G014 (Fabbiano et al 2018c), two knots of Fe Kα emission are detected, with ~30 pc projected separation, embedded in a larger diffuse region of both hard (3-6 keV) continuum and Fe Kα emission, extending out to 1-2 kpc in the direction of the ionization bicone (Fabbiano et al 2017, 2018a, b). These knots are inclined by ~30 degrees to the bi-cone axis, rather than being at ~90 degrees, as in NGC 4945. In NGC 5643, discussed here, the ~65 pc N-S feature we have detected is perpendicular to the ionization cone that extends E-W (Morris et al 1985; Schmitt et al 1994; Cresci et al 2015), as in NGC 4945.

The details of the morphology of these nuclear regions are somewhat energy dependent in all cases. In particular, in NGC 4945 (Marinucci et al 2017) the equivalent width (EW) of the Fe Kα line is spatially variable (ranging from 0.5 to 3 keV), on scales of tens of parsecs. This EW spatial variability suggests differences in the ionization state and spatial distribution of the optically thick reprocessing clouds, with respect to the central X-ray illuminating source. In ESO 428-G014 (Fabbiano et al 2018c), we reported EW of ~1 and > 2 keV for the nuclear knots.

In NGC 5643, using the regions in Fig. 3, and comparing the 6.2 – 6.6 keV with the 5.0-6.0 keV continuum images, we find values of EW consistent with those of NGC 4945 and ESO 428-G014. In particular, EW = 1.6±0.5 keV (north), 1.4±0.3 keV (center), and 2.5± 0.8 keV (south). The total Fe Kα EW using counts from a box encompassing the N-S emission is 1.6±0.2. As argued for ESO 428-G014 (Fabbiano et al 2017), these large EW values exclude that the N-S feature may be due to luminous X-ray binaries. The lack of a corresponding N-S structure in the 3-6 keV continuum reinforces this conclusion and identifies the N-S structure with reflecting CT circumnuclear clouds.

3.2 The Fe Kα EW of CT AGNs

The EW gives a measure of the reprocessing of the continuum to generate the Fe Kα line. The Fe Kα EW is typically large in highly obscured CT AGNs. As discussed in Section 3.1 the empirical EW derived from *Chandra* data by comparing the Fe Kα emission with the ~4-6 keV continuum are all in the ~1-2 keV range, and they show local variations.

The discovery that much of the Fe Kα emission arises at large distances and has a wide range of EW (see Fabbiano et al 2017 and Section 3.1), complicates the use of this EW as a diagnostic tool for AGN evolution. In particular, Boorman et al. (2018) claim an anticorrelation between Fe Kα EW and continuum luminosity, as measured by the 12μm luminosity (Asmus et al. 2016). With a 12μm luminosity of ~$10^{43}$ erg s$^{-1}$ (NED, Sanders et al., 2003) each of the North, center, and South regions of the N-S feature of NGC 5643 lie above the best-fit correlation in Boorman et al. (2018).

3.3 Imaging the Torus

The idea of AGN obscured by a CT torus (the standard model, see e.g. Antonucci 1993), is generally accepted, since it manages to explain several observational characteristics of AGNs. However, it is only recently that we begin to have direct observational evidence of these structures.

As the maximum dust sublimation temperature at the nucleus of NGC 4945 (Asmus et al. 2015) implies an inner radius of ~8 pc for the AGN torus, Fabbiano et al (2018c) argue that the ~10 times larger flattened nuclear distribution of Fe Kα in NGC 4945 may arise from scattering off molecular clouds in a flattened structure in the ISM of NGC 4945. Similar large-scale molecular gas is common in other AGN, based on mapping of the $H_2$ (2.12 μm) emission line (Hicks et al., 2009) who argue that this flattened larger structure is connected with the smaller scale AGN torus based on their co-alignment.

In NGC 5643, the ALMA detection and kinematics of a rotating circumnuclear molecular disk indicate that this disk is unconnected with the general galaxy ISM. The *Chandra* detection of a corresponding N-S structure in the neutral Fe Kα line, would argue for both CO and Fe Kα emission originating from the obscuring torus. The ALMA position-velocity diagram shows that, kinematically, the CO disk extends to +/-~0.3", but not to 0.4" (Alonso-Herrera et al. 2018, Fig.8.). If the entire N-S feature were due to Fe Kα fluorescence off an optically thick scattering disk, this disk would be larger than the ALMA CO disk. However, if we assume that the most luminous clump is coincident with the nucleus, the Fe Kα emission could have the same size of the CO disk, if the northernmost clump is not part of this dynamical structure (fig. 5).

The ALMA CO disk has a line of sight $N_H$ ~few x $10^{23}$ cm$^{-3}$, (Alonso-Herrera et al. 2018, Sec.4.4). This is ~10 times lower than the minimum Compton Thick $N_H$, as seen in the X-ray spectrum (Risaliti et al 1999). Assuming a similar $N_H$ radially from the AGN continuum source then, as in the Hicks et al. (2009) $H_2$ study, this discrepancy implies a clumpy disk, although as the X-ray nucleus provides only a single point, the requirement is less strong in NGC 5643. In NGC 5643 clear lines of sight ($N_H \sim < 10^{23}$ cm$^{-2}$) through the torus are needed to allow the X-ray continuum through to the outer molecular clouds where it can create the fluorescent Fe K$\alpha$ line, i.e. a clumpy torus (Nenkova et al., 2008) with a low density (<$n_e$> $\sim < 10^3$ cm$^{-2}$) inter-cloud medium, in contrast to some torus models (Stalevski et al. 2016). Hicks et al. (2009) also require a clumpy disk.

The ALMA map is slightly asymmetric in CO(2-1) intensity, from 2.0 Jy km s$^{-1}$ beam$^{-1}$ in the North to 2.4 Jy km s$^{-1}$ beam$^{-1}$ in the South (Alonso-Herrera et al. 2018, their figure 5). If we take this as a measure of relative $N_H$, then the Fe-K emission will be stronger in the South by factor $e^{-2.0}/e^{-2.4}$ = 1.5. If we use the 1/16 image of Fig. 5 as guidance, and we exclude the northernmost clump (see above), we find that the ratio between the clump south of the nucleus appears brighter than the one just north of the nucleus. However, the counting statistics is poor, leading to a ratio of 1.7±0.7. Comparison of the northernmost and southernmost clumps shows that the southern clump yields slightly more counts, but the two values are compatible within the statistical error.

In NGC 4945 Marinucci et al (2017) finds also an extended clumpy Fe XXV nuclear feature, with some spatial difference from the neutral Fe K$\alpha$ structure, which could highlight regions of prevalent shock excitation. In NGC 5643 the 6.6-6.9 keV emission shows a single peak, consistent with the continuum peak.

## 4. Summary and Conclusions

In summary:

1. Imaging the deep on-axis *Chandra* ACIS observations of the CT AGN NGC 5643 in the 6.2-6.6 keV Fe K$\alpha$ line, we have found a significant elongated feature in the nuclear region. This feature is clumpy and extends for ~65 pc N-S. No corresponding feature is seen in the 3.0-6.0 keV continuum.
2. We find EW of 1.6±0.5 keV (north), 1.4±0.3 keV (center), and 2.5± 0.8 keV (south). The total Fe K$\alpha$ EW using counts from a box encompassing the N-S emission is 1.6±0.2
3. The Fe K$\alpha$ feature is spatially consistent with the N-S elongation found in the CO(2-1) high resolution imaging with *ALMA* ( Alonso-Herrero et al 2018), but slightly more extended than the rotating molecular disk of r=26 pc indicated by the kinematics of the CO(2-1) line.
4. The *Chandra* detection of a corresponding N-S structure in the neutral Fe K$\alpha$ line, would argue for both CO and Fe K$\alpha$ emission originating from the obscuring torus.
5. Both the clumpiness of the Fe K$\alpha$ map and the relatively low $N_H$ ~few x 10$^{23}$ cm$^{-3}$, derived from the *ALMA* data (Alonso-Herrera et al. 2018, Sec.4.4), suggest a clumpy obscuring circumnuclear disk.

In conclusion, both high resolution *Chandra* X-ray data and ALMA observations in the sub-mm range are beginning to provide direct pictures of the circumnuclear regions of CT AGNs. These observations are complementary in that the hard X-ray and Fe Kα emission is due to AGN photons scattered by circumnuclear CT clouds, while the molecular clouds themselves are directly revealed via their CO emission with ALMA. NGC 5643 – discussed here-, is the first instance of a CT AGN with both Fe Kα and CO(2-1) observation at scales <100 pc where both windows suggest a N-S CT feature, that could be connected with the 'torus' of the standard AGN model.

A full study of the X-ray emission of NGC 5643, including spectral modeling, will be presented in a forthcoming paper.

Larger scale *Chandra* imaging in the hard X-ray continuum and Fe Kα line has also shown that these emissions are not just confined to the immediate circumnuclear region. They can arise at radii as large as a few kiloparsecs. These features suggest interaction with dense molecular clouds in the host galaxy disk (Fabbiano et al 2017, 2018a, 2018b), opening the way for future joint *Chandra – ALMA* studies. These exploratory deep *Chandra* observations argue for the need for a large area, high-resolution future X-ray observatory, such as *Lynx*, to pursue observational studies of AGN-galaxy interaction in detail in a large number of AGNs.

We retrieved data from the NASA-IPAC Extragalactic Database (NED) and the Chandra Data Archive. For the data analysis, we used the CIAO toolbox and DS9, developed by the Chandra X-ray Center (CXC). This work was partially supported by the Chandra Guest Observer program grant GO5-16090X (PI: Fabbiano) and NASA contract NAS8-03060 (CXC).

**REFERENCES**

Alonso-Herrero, A., Pereira-Santaella, M., Garcia-Burillo, S. et al 2018, ApJ, 859, 144.

Annuar, A., Gandhi, P., Alexander, D. M. et al 2015, ApJ, 815, 36.

Antonucci, R., 1993, Ann. Rev. Astron. Astrophys., 31, 473.

Asmus, D., Gandhi, P., Höning, S. F., Smette, A. and Duschl, W. J. 2015, MNRAS, 454, 766.

Boorman, P. G., Gandhi, P., Baloković, M., Brightnan, M., Harrison, F., Ricci, C., and Stern, D. 2018, MNRAS, 477, 3775.

Cresci, G., Marconi, A., Zibetti, S. et al 2015, A&A, 582, 63.

Fabbiano, G., Elvis, M., Paggi, A., Karovska, M., Maksym, W. P., Raymond, J., Risaliti, G., and Wang, Junfeng, 2017, ApJL, 842L, 4.

Fabbiano, G., Paggi, A., Karovska, M., Elvis, M., Maksym, W. P., Risaliti, G., and Wang, Junfeng 2018a, ApJ, 855, 131.

Fabbiano, G., Paggi, A., Karovska, M., Elvis, M., Maksym, W. P., Risaliti, G., and Wang, Junfeng 2018b, ApJ, 865, 83.

Fabbiano, G., Siemiginowska, A., Paggi, A., Elvis, M., Volonteri, M., Mayer, L., Karovska, M., Maksym, W. P., Risaliti, G., and Wang, Junfeng 2018c, ApJ in press, http://arxiv.org/abs/1811.06436

Fruscione, A., McDowell, J. C., Allen, G. E. et al 2006, SPIE Proc., Vol. 6270, id. 62701V

Guainazzi, M., Rodriguez-Pascual, P., Fabian, A. C., Iwasawa, K. and Matt, G. 2004, MNRAS, 355, 297.

Hicks, E. K. S., Davies, R. I., Malkan, M. A., Genzel, R., Tacconi, L. J., Müller Sánchez, F. and Sternberg, A. 2009, ApJ, 696, 448.

Levenson, N. A., Heckman, T. M., Krolik, J. H., Weaver, K. A., and Życki, P. T. 2006, ApJ, 648, 111.

Matt, G., Bianchi, S., Marinucci, A., Guainazzi, M., Iwawasa, K. and Jimenez Bailon, E. 2013, A&A, 556, 91.

Marinucci, A., Risaliti, G., Wang, J., Nardini, E., Elvis, M., Fabbiano, G., Bianchi, S. and Matt, G. 2012, MNRAS, 423, L6.

Marinucci, A., Bianchi, S., Fabbiano, G., Matt, G., Risaliti, G., Nardini, E., and Wang, J. 2017, MNRAS, 470, 4039.

Morris, S. L., Ward, M. J., Whittle, M., Wilson, A. S. and Taylor, K. 1985, MNRAS, 216, 193.

Nenkova, M., Sirocky, M.M., Nikutta, R., Ivezic, Z., and Elitzur, M., 2008, ApJ, 685, 160.

Risaliti, G., Maiolino, and R. Salvati, M. 1999, ApJ. 522, 157.

Schmitt, H. R., Storchi-Bergmann, T. and Baldwin, J. A. 1994, ApJ, 423, 237.

Stalevski, M., Ricci, C., Ueda, Y., Lira, P., Fritz, J. and Baes, M. 2016, MNRAS, 458, 2288.